\documentclass[12pt]{article} 

\begin{document}
\newcommand{\be}{\begin{equation}}
\newcommand{\ee}{\end{equation}}
\newcommand{\bea}{\begin{eqnarray}}
\newcommand{\eea}{\end{eqnarray}}
\pagestyle{empty}

\begin{center}
{\LARGE\bf Yangians, finite $\cal W$-algebras\\
 and the\\[2.1ex]
Non Linear Schr{\"o}dinger hierarchy}
\\[2.1em]
\end{center}

{\Large
M. Mintchev$^{a}$\footnote{mintchev@difi.unipi.it},
E. Ragoucy$^{b}$\footnote{ragoucy@lapp.in2p3.fr}, 
P. Sorba$^{b}$\footnote{sorba@lapp.in2p3.fr} and 
Ph. Zaugg$^{c}$\footnote{zaugg@CRM.UMontreal.CA}}\\

\indent

\noindent
{\it $^a$ INFN, Dipart. di Fisica dell'Univ. di
  Pisa, Piazza Torricelli 2, 56100 Pisa, Italy\\[2.1ex]
$^b$ LAPTH, Chemin de Bellevue, BP 110, F-74941 Annecy-le-Vieux
  cedex, France\\[2.1ex]
$^c$ CRM, Univ. de Montr{\'e}al, CP 6128, 
Succ. Centre-ville, Montr{\'e}al, 
Qc H3C 3J7 Canada}

\vfill

\begin{center}
Talk presented by E. Ragoucy at the ACTP-Nankai Symposium\\
 {\it
  Yang-Baxter systems, non linear models and their 
applications}\\
  Seoul (Korea) October 20-23, 1998.
\end{center}

\vfill

\begin{abstract}
We show an algebra morphism between Yangians and some finite 
$\cal W$-algebras. This correspondence is nicely illustrated in the
framework of the Non Linear Schr{\"o}dinger hierarchy. For such a
purpose, we give an explicit realization of the Yangian generators in
terms of deformed oscillators.
\end{abstract}

\vfill
\rightline{LAPTH-711/98-Conf}
\newpage
\pagestyle{plain}
\setcounter{page}{1}
\section{Introduction}

\indent

${\cal W}$ algebras
first showed up in the context of two
dimensional conformal field theories\cite{1}. 
Yangians were first considered and defined in
connection with some rational solutions of the quantum Yang-Baxter
equation\cite{3}. 
 
In this note, we will show that the defining relations of a Yangian
are satisfied for a family of finite ${\cal W}$ algebras (FWA). 
In other words, such ${\cal W}$-algebras
provide Yangian realizations. This remarkable connection between two a
priori different types of symmetry deserves in our opinion to be
considered more thoroughly. We have already shown how this
 result can help for the classification of the
irreducible finite dimensional representations of 
FWA's\cite{13}. 
Here, we show
that the Non Linear Schr{\"o}dinger Hierarchy is a nice framework where the connection
is visualised. For such a purpose, we first study this hierarchy and
construct the Yangian generators using a deformed oscillator algebra. 

This report is a condensed version of \cite{13,MRSZ}.

\section{Finite ${\cal W}(sl(nm), n.sl(m))$ algebras}

The usual notation for a ${\cal W}$ algebra obtained by the Hamiltonian
reduction procedure is ${\cal W}({\cal G}, {\cal H})$\cite{2,16}. More precisely, given a
simple Lie algebra ${\cal G}$, there is a one-to-one correspondence between
the finite ${\cal W}$ algebras one can construct in ${\cal U}({\cal G})$ and the $sl(2)$
subalgebras in ${\cal G}$. Since any
$sl(2) \ {\cal G}$-subalgebra is
principal in a subalgebra ${\cal H}$ of
${\cal G}$, it is rather usual to denote the corresponding ${\cal W}$
algebra as ${\cal W}({\cal G}, {\cal H})$.

\indent

As an example\cite{9}, let us consider the ${\cal W}(sl(4), 2.sl(2))$
algebra, where $2.sl(2)$ stands for $sl(2)\oplus sl(2)$. 
It is made of seven generators $J_i, S_i\ (i=1,2,3)$ 
and a central
element $C_2$ such that:
\begin{equation}
\begin{array}{ll}
  {[}J_i, J_j] = i {\epsilon_{ij}}^k J_k & (i,j,k=1,2,3)  \\
  {[} J_i, S_j ] = i {\epsilon _{ij}}^k S_k &  \\
  {[} S_i, S_j] = -i {\epsilon _{ij}}^k J_k (2 \vec{J}^2 - C_2 - 4)& \\
  {[}C_2, J_i ] = [C_2 , S_i] =0 & \mbox{with } \vec{J}^2 =
  J^2_1 + J^2_2 + J_3^2
  \label{1}
\end{array}
  \label{eq:1}
\end{equation}

We recognize the $sl(2)$ subalgebra generated by the $J_i$'s as well
as an adjoint representation (i.e. $S_i$ generators) of this $sl(2)$
algebra. We note that the $S_i$'s close polynomially on the other
generators. If one assumes a ``degree'' 1 for the $J$'s and 2 for the
$S$'s, we remark that the commutator of a degree $k$ generator with
a degree $j$ one is of degree $j+k-1$.

\indent

The same type of
structure can be remarked, at a higher level, for the class of
algebras ${\cal W}(sl(nm),n.sl(m))$. This algebra is formed by
generators $W_k^a$ ($a=1,\dots, (n^2-1)$ and $k=1,\dots,m$) together
with central elements $C_k$. The $W$ generators gather into
 a stack of $m$
adjoint representations of $sl(n)$ (the first one being an $sl(n)$
algebra), indexed by the degree $k$, and which
close polynomially, respecting the degree: $[W^a_k, W^b_j]=
P^{ab}_{k+j-1}$ where $P$ is a polynomial in $W$ and $C$
generators. 

\section{Yangians $Y({\cal G})$}

Yangians are  infinite dimensional 
quantum groups\cite{3,15}
 that correspond to 
quantization of (half of) the loop
algebra of some finite-dimensional Lie algebra ${\cal G}$. 
As such, it is a Hopf algebra, possesses a R-matrix and can be defined
through a $RTT=TTR$ relation. 
Here, we will choose an alternative
approach (known to be equivalent) that 
enlights 
the loop algebra deformation and which is best suited to our
purpose. In that context, the Yangian $Y({\cal G})$ is generated by an
infinite stack of adjoint representations of ${\cal G}$, indexed by a
degree $n$ going from 0 to infinity, the degree 0 corresponding to
${\cal G}$ itself, the generators  of degree 1 being subject to the
following constraints:
\[
\begin{array}{l}
 {f^{bc}}_d {[Q_1^a,Q_1^d]} + {f^{ca}}_d {[Q_1^b,Q_1^d]} 
+ {f^{ab}}_d {[Q_1^c,Q_1^d]} =
 {f^a}_{pd}{f^b}_{qx}{f^c}_{ry}{f^{xy}}_e
\eta^{de}\ s_3(Q_0^p,Q_0^q,Q_0^r) \\
 {f^{cd}}_e {[[Q_1^a,Q_1^b],Q_1^e]} + 
{f^{ab}}_e {[[Q_1^c,Q_1^d],Q_1^e]}\  =\   \\
 \ \ \ \ \left(
{f^a}_{pe}{f^b}_{qx}{f^{cd}}_y{f^y}_{rz}{f^{xz}}_g+
{f^c}_{pe}{f^d}_{qx}{f^{ab}}_y{f^y}_{rz}{f^{xz}}_g\right)
 \eta^{eg}\ s_3(Q_0^p,Q_0^q,Q_1^r) \nonumber
\end{array}
\]
where ${f^{ab}}_c$ are the totally antisymmetric structure constant of 
${\cal G}$, 
$\eta^{ab}$ is the Killing form, and $s_3(.,.,.)$ is the 
totally symmetrized
product of $3$ terms.
It can be shown that for ${\cal G}=sl(2)$, the first constraint 
is trivially satisfied, while for ${\cal G}\neq sl(2)$, the
last constraint follows from the previous one. 

Note that the constraints imply only the $Q_0$ and $Q_1$ generators,
 $Y({\cal G})$ being totally defined once we have said that it is a
 homogeneous quantization of the loop algebra on ${\cal G}$.

In the following, we will focus on the Yangians $Y(sl(n))$.

\section{${\cal W}(sl(nm), n.sl(m))$ as a realisation of $Y(sl(n))$.}
{F}rom the previous definitions of ${\cal W}(sl(nm), n.sl(m))$ 
algebras and Yangian $Y(sl(n))$, it is natural to look for a relation
between these objects. Indeed, it can be proven that\cite{13}:

{\em Identifying the generators $W^a_k$ of the finite ${\cal W}(sl(nm), n.sl(m))$
algebra with the elements $Q^a_{k-1}$ ($k=1,\dots m$) of the Yangian $Y(sl(n))$, one
verifies that the defining relations of a Yangian are satisfied for
this ${\cal W}$ algebra, which therefore appears as a realisation of the Yangian $Y(sl(n))$.}

Owing to the above identification, it is possible to link the
irreducible finite dimensional representations of the Yangians 
(known to be products of
 evaluation representations) to the Miura map developed in the
framework of ${\cal W}$ algebras. We will not develop these points, inviting the
interested reader to look at \cite{13,14} for more information. 
Instead, we will focus on a ``physical'' framework where the
identification can be ``visualised''.

\section{Non Linear Schr{\"o}dinger equation and its hierarchy}
\subsection{State of the art}

We start with the well-known Non Linear Schr{\"o}dinger equation 
(NLS) in 1+1 dimension, on the real line:
\be
i\frac{\partial\phi(x,t)}{\partial t} +
\frac{\partial^2\phi}{\partial^2 x} = 2g |\phi|^2\ \phi
\ee
where $\phi$ is a complex field. 

In the following we will look at the ``vectorial'' version of it:
\be
i\frac{\partial\phi(x,t)}{\partial t} +
\frac{\partial^2\phi}{\partial^2 x} = 2g\ (\phi^\dag\cdot\phi) \phi
\mbox{ with } 
\phi=\left(\begin{array}{c}\varphi_1 \\ \varphi_2 \\
\vdots \\ \varphi_N\end{array}\right)
\ee

The solution to this equation has
being given (at the classical level) by Rosales\cite{Ros}:
\bea
\phi(x,t) &=& \sum_{n=0}^\infty (-g)^n \phi_n(x,t) \
\mbox{ where} \label{phi:g}\\
\phi_n &=& \int d^{n+1}q\, d^np\ \frac{\bar\lambda(p_1) 
\bar\lambda(p_2)\cdots\bar\lambda(p_n)
\lambda(q_n) \cdots\lambda(q_1)\lambda(q_0)}
{\prod_{j=1}^n(p_j-q_j)(p_j-q_{j-1})}
\exp\left(i\Omega_n\right)\nonumber\\
\mbox{with} && \Omega_n=\sum_{j=0}^n (xq_j-tq^2_j)-
\sum_{j=1}^n (xp_j-tp^2_j)
\eea
with $\lambda$ an arbitrary function (such that
the series is well-defined). A remarkable
fact is that this solution is also valid at the quantum level\cite{solQ}, or more
precisely if one looks for a local field $\phi$ satisfying
\bea
&& {[\phi(x,t),\phi^\dag(y,t)]}=\delta(x-y) \mbox{ and }
{[\phi(x,t),\phi(y,t)]}=0 \ \mbox{ where:}\nonumber\\
&& \phi(x,t)=e^{iHt}\phi(x,0)e^{-iHt}
 \mbox{ with }
H=\int dx\ \frac{\partial\phi^\dag}{\partial x}\cdot
\frac{\partial\phi}{\partial x} + g (\phi^\dag\cdot \phi)^2
\label{H:phi}
\eea
then, the solution is given by a series
expansion of the above type with
\be
\phi_n = \sqrt{2\pi}\int d^{n+1}q\, d^np\ \frac{a^\dag(p_1) 
a^\dag(p_2)\cdots a^\dag(p_n)
a(q_n) \cdots a(q_1)a(q_0)}
{\prod_{j=1}^n(p_j-q_j-i\epsilon)(p_j-q_{j-1}-i\epsilon)}\ 
e^{i\Omega_n}
\label{phi:a}
\ee
where now $a(p)$ and $a^\dag(p)$ generate a deformed oscillator algebra
(or ZF algebra \cite{ZF}):
\bea
a_\alpha(p)a_\beta(p') &=& R_{\beta\alpha}^{\nu\mu}(p'-p)
a_\mu(p')a_\nu(p)\\
{a^\dag}^\alpha(p){a^\dag}^\beta(p') &=&
{a^\dag}^\mu(p'){a^\dag}^\nu(p)
R^{\alpha\beta}_{\mu\nu}(p'-p)\\
a_\alpha(p){a^\dag}^\beta(p') &=& 
{a^\dag}^\mu(p')R_{\alpha\mu}^{\beta\nu}(p-p')
a_\nu(p)+\delta_\alpha^\beta\delta(p-p')
\label{a*a}
\eea
The construction (\ref{phi:a}) is understood on the Fock space of this
ZF algebra \cite{fock}.
Note that the indices $\alpha$, $\beta,\dots$ run from 1
to $N$, because the vector $\phi$ is in the fundamental
representation of $sl(N)$.
The R-matrix appearing in this algebra is just the one of 
$Y(sl(N))$.
This is not surprising, since $Y(sl(N))$ is a symmetry of NLS. In
fact, it has been shown in \cite{Wad} that, for $N=2$, the 
Yangian generators
can be expressed in term of $\phi$ and Pauli matrices $t^a$:
\bea
J^a &=& \int dx\, \phi^\dag(x)t^a\phi(x)\label{Q:phi}\\
S^a &=& \! \frac{i}{2}\int dx\, \phi^\dag(x)t^a\partial_x\phi(x)
-\frac{ig}{2}\int dxdy\ sgn(y-x)
\left(\phi^\dag(x)t^a\phi(y)\right)
\phi^\dag(x)\cdot\phi(y)\nonumber
\eea
where $J^a$ stands for the generator $Q^a_0$, and 
$S^a$ for $Q^a_1$.
It is easy to see that this formula is also valid in the general
case $Y(sl(N))$, the generators $t^a$ being in the fundamental
representation of $sl(N)$.

It is natural to look for the form of the Yangian generators
in terms of the ZF generators. We are going to see that the answer,
although not so trivial, will reveal the natural link between
Yangians and ${\cal W}$-algebras, and will also lead to a very nice
formulation of the NLS hierarchy.

\subsection{NLS hierarchy}
It is interesting to rewrite 
the Hamiltonian (\ref{H:phi}), the total momentum $P=-i\int dx
\phi^\dag\cdot \partial_x\phi$ and the particle number $N=\int dx 
\phi^\dag\cdot \phi$ in term of the ZF algebra. The result is
surprisingly simple:
\bea
N &=& \int dp\ a^\dag(p)\cdot a(p)\ \ ;\ \ 
P \ =\  \int dp\ p\ a^\dag(p)\cdot a(p)\\
H &=& \int dp\ p^2\ a^\dag(p)\cdot a(p)
\eea
More generally, one can consider the operators
\be
H_k = \int dp\ p^k\ a^\dag(p)\cdot a(p)
\ee
A very nice result is \cite{MRSZ}:

{\em i) $H_k$ is the Hamiltonian of the $k^{th}$ equation of
  the NLS hierarchy. 

ii) The local field evolving with $H_k$ takes the form
(\ref{phi:g}), (\ref{phi:a}), with now}
\be
 \Omega_{k,n}=\sum_{j=0}^n (xq_j-tq^k_j)-
\sum_{j=1}^n (xp_j-tp^k_j)
\ee

\subsection{Yangian and deformed oscillators}
In view of the formulae (\ref{phi:a}) and (\ref{Q:phi}), it is clear that trying
to reconstruct the Yangian generators (in term of oscillators) by
direct calculation is a difficult task. Fortunately, the commutation
relations of the oscillators with these generators have being 
given in \cite{Wad}, and it is simpler to seek for operators that
fulfill these relations (see \cite{MRSZ} for details). 
To simplify the presentation, we drop the indices $\alpha$,
$\beta,\dots$ and the momenta, and use indices refering to
 spaces instead. For instance, (\ref{a*a}) reads 
$a_1a^\dag_2=a^\dag_2R_{21}a_1+\delta_{12}$.
{F}or $sl(2)$, a careful calculation leads to (the general 
case can be found in \cite{MRSZ}):
\bea
J^a &=& \sum_{n=1}^\infty \frac{(-)^{n+1}}{n!} J^a_n 
\ \ \mbox{ and }\ \ 
S^a \ =\ \sum_{n=1}^\infty \frac{(-)^{n+1}}{n!} 
S_n^a \label{Q:a}\\
J^a_n &=&  a^\dag_{1\dots n} T^a_{1\dots n} a_{n\dots 1} 
\ \ \mbox{ with }T^a_{1\dots n}= \sum_{j=1}^n (-)^{j-1}
\left(\begin{array}{c} n-1\\j-1 \end{array}\right) t^a_j
 = \sum_{j=1}^n \alpha^n_j\, t^a_j \nonumber\\
S^a_n &=&  a^\dag_{1\dots n} \widetilde{T}^a_{1\dots n} 
a_{n\dots 1} \ \ \mbox{ with }
\widetilde{T}^a_{1\dots n}=\sum_{j=1}^n \alpha^n_j\,\left(
p_j\, t^a_j\, -ig{f^{a}}_{bc}\, 
\sum_{i=1}^j  t^{b}_i t^c_j\right)\nonumber
\eea
where 
$a^\dag_{1\dots n}=a^\dag_1(p_1)a^\dag_2(p_2)\cdots a^\dag_n(p_n)$,
$a_{n\dots 1}=a_n(p_n)a_{n-1}(p_{n-1})\cdots a_1(p_1)$ 
and the 
integration on
$p_1, p_2,\dots,p_n$ is implied in $J^a$ and $S^a$. 

The formulae 
(\ref{Q:a}) provide a construction for the Yangian generators
in terms of the ZF algebra.
In this formulation, 
it
is easy to see that the Yangian generators commute with $H_k$, so that
the Yangian symmetry of the whole NLS hierarchy is manifest. 

\subsection{${\cal W}$-algebra in NLS hierarchy}
Let us now focus on the Fock space ${\cal F}$ 
associated to the ZF algebra. It
can be decomposed into a direct sum of subspaces with fixed particle
number: ${\cal F}=\oplus_{m=0}^\infty {\cal F}_m$. Since the
Yangian generators commute with the particle number, 
we can consider their restriction to any ${\cal F}_m$. 
On that subspace,
the infinite series (\ref{Q:a}) truncate, because the products of more
than $m$ $a$'s identically vanish on  ${\cal F}_m$.
We are thus considering polynomials in $a$'s and $a^\dag$'s of order
less than
$m+1$. This implies that there is only a finite number of independent Yangian generators
(considered as operators on ${\cal F}_m$), the other ones being in their
enveloping algebra. In other words, we get a
polynomial algebra which satisfies the defining relation of the
Yangian $Y(sl(N))$: it is
related to
the ${\cal W}(sl(mN),Nsl(m))$ algebra. More precisely, the
Hamiltonians $H_k$ ($k\leq m$) have to be added and generate 
a $m$-dimensional center of this polynomial algebra,
 while the ${\cal W}(sl(mN),Nsl(m))$
algebra possesses only $m-1$ central generators. One thus needs a
constraint to connect these two algebras. It appears to be $H_1=0$,
{\it i.e.} the vanishing of the total momentum. On the corresponding
Fock space ${\cal F}^{(red)}_m$, the Yangian 
will be represented by a ${\cal W}$-algebra. Note
that this constraint is just the one introduced in \cite{13,14} when looking
at Yangian evaluation representations as ${\cal W}$ representations
(see \cite{MRSZ} for details).

In the special case of $m=1$,
since $S^a_1=0$ (on ${\cal F}^{(red)}_1$), 
we recover a 
$sl(N)$ algebra. 

\section{Conclusion}

There is a natural correspondence between ${\cal W}(sl(mn),n.sl(m))$
algebras and Yangians $Y(sl(n))$. It is well illustrated in the
framework of NLS hierarchy, where the use of a deformed oscillator
algebra makes explicit the Yangian symmetry of this hierarchy. On
the Fock space of the ZF algebra, the action of the Yangian is represented
(for fixed particle number $m$ and vanishing total momentum) by a ${\cal W}(sl(mn),n.sl(m))$
algebra. This is visualised in the figure \ref{YWF}.
\begin{figure}[htb]
\[
\begin{array}{ccccccc}
 & & \vdots & & \vdots & & \\
& & W(sl(m'N),N.sl(m')) & >>> & {\cal F}^{(red)}_{m'} & & \\
 &\nearrow &\vdots &  &\vdots & \searrow & \\
Y(sl(N)) & \rightarrow & W(sl(mN),N.sl(m))& >>> 
& {\cal F}^{(red)}_m & \rightarrow & {\cal F}^{(red)} \\
 &\searrow & \vdots & & \vdots &\nearrow & \\
 & & W(sl(2N),N.sl(2)) & >>> & {\cal F}^{(red)}_2 & & \\
 & & sl(N) & >>> 
& {\cal F}_1^{(red)} & & \\
\end{array}
\]
\caption{{\bf Action of the Yangian on the reduced Fock space.}
 {\it On
  each subspace ${\cal F}^{(red)}_m$ (fixed particle number $m$ and
  vanishing total momentum), it is realized by a finite
  ${\cal W}$ algebra which acts on ${\cal F}^{(red)}_m$.}
\label{YWF}}
\end{figure}

This construction can be generalised to the case of NLS on a
half-line. In that case, Yangians are replaced by 
twisted
 Yangians
\cite{Ytwist}, the ZF algebra by an enlarged deformed oscillator
algebra which takes into account the boundary properties \cite{mih},
and the ${\cal W}$ algebra by a folded ${\cal W}$ algebra
\cite{fold}. This work is under investigation \cite{MRSZ2}.

\end{document}